\documentclass[twocolumn,aps,prb,showpacs]{revtex4}
\usepackage{graphicx}

\begin{document}

%
%

\title{Landau level transitions in InAs/AlSb/GaSb quantum wells}

\author{X. G. Wu and M. Pang}

\affiliation{SKLSM, Institute of Semiconductors, Chinese Academy
of Sciences, Beijing 100083, China}

\begin{abstract}

The electronic structure of InAs/AlSb/GaSb quantum wells embedded
in AlSb barriers and in the presence of a perpendicular magnetic
field is studied theoretically within the $14$-band ${\bf k}\cdot{\bf p}$
approach without making the axial approximation.
At zero magnetic field, for a quantum well with a wide InAs layer and
a wide GaSb layer, the energy of an electron-like subband can be lower than
the energy of hole-like subbands.  As the strength of the magnetic
field increases, the Landau levels of this electron-like subband
grow in energy and intersect the Landau levels of the hole-like
subbands.  The electron-hole hybridization leads to a series of
anti-crossing splittings of the Landau levels.
The energies of some Landau level transitions and their corresponding
transition strengthes are calculated.  The magnetic field dependence
of some dominant transitions is shown with their corresponding initial-states
and final-states indicated.  This information should be useful in
analyzing an experimentally measured magneto-optical spectrum.
At high magnetic fields, multiple transitions due to the initial-state
splitting can be observed.  The dominant transitions at high fields can be
roughly viewed as two spin-split Landau level transitions with
many electron-hole hybridization induced splittings.
The energy separations between the dominant transitions may decrease or
increase versus the magnetic field locally, or may be almost field
independent.  The separations can be tuned by changing the width
of InAs layer or the width of middle AlSb layer.
When the magnetic field is tilted, the electron-like Landau level
transitions show additional anti-crossing splittings due to the
subband-Landau level coupling.

\end{abstract}


\maketitle

%
%

\section{Introduction}

In the past decades, the transport and optical properties of
InAs/GaSb quantum wells have been investigated experimentally and
many interesting features have been revealed.\cite{exp01,
exp02, exp03, exp04, exp05, exp06, exp07, exp08, exp09, exp10, exp11, top03, top04}
Among various experimental techniques, the Landau level spectroscopy
has been employed in exploring the electronic properties of InAs/GaSb
quantum wells.  A key element in understanding those experiments is
the electron-hole hybridization.\cite{theory01}
At low magnetic fields, the agreement between the
experiments and theories seems to be fairly good.\cite{theory02}
However, at high magnetic fields, the difference between the
experimental data and the theoretical simulation is small for some
quantum structures, but is not small for some other structures.\cite{exp10}
A dominate transition is observed at high magnetic fields, but
this is unexpected as the corresponding initial state should be empty.\cite{exp10}
In a more recent experimental study, without a direct comparison
to a theoretical simulation, it is pointed out that the
conventional model can not account for the observed features, and
one needs to introduce a spontaneous phase separation for the
nominally uniform InAs/AlSb/GaSb quantum wells.\cite{exp11}

Most previous theoretical works on the Landau levels transitions
in InAs/AlSb/GaSb quantum wells are based on the six-band or eight-band
${\bf k}\cdot{\bf p}$ approach.\cite{theory02, theory03, theory04}
Usually, the transition energy is given but the transition strength
is not shown.  Furthermore, the axial approximation is often used in these
theoretical calculations.  When the
axial approximation is employed, some couplings between Landau levels
are neglected, and the influence of bulk inversion asymmetry, which
is present in the system, is usually ignored.  For HgTe quantum wells,
where the electron-hole coupling is also important, it is
necessary to avoid these approximations in order to understand some
experimental results.\cite{hgte01, hgte02, hgte03}  Another potential
problem is the possibility of spurious solution.\cite{sp01, sp02}
It is known to occur in the calculation of electronic states in
some InAs/GaSb superlattices within the eight-band ${\bf k}\cdot{\bf p}$
approach.  Thus, it is better for one to go beyond the eight-band
${\bf k}\cdot{\bf p}$ method in the study of InAs/AlSb/GaSb quantum
wells.

In the present paper, we investigate theoretically the electronic
structure of InAs/AlSb/GaSb quantum wells embedded in AlSb barriers
and in the presence of a perpendicular magnetic field within
the $14$-band ${\bf k}\cdot{\bf p}$ approach without making the
axial approximation.  The energies of some Landau level transitions
and their corresponding transition strengthes are calculated.
The magnetic field dependence of some dominant transitions is displayed
together with their corresponding initial-states and final-states
indicated.  This information should be useful in analyzing an experimentally
measured magneto-optical spectrum.  Unfortunately, a
quantitative comparison with the experiments is not yet possible, as
the distribution of donors and acceptors in the experimental samples
is not known.  The knowledge of this distribution is necessary in determining the
magnetic field dependence of the self-consistent potential.  This
self-consistent potential affects the relative position of the electron-like
and the hole-like subands and hence the features shown in the magnetic field
dependence of the Landau level transitions.  We wish that the present
theoretical work will inspire more experimental investigations.

This paper is organized as follows: In section II, the theoretical
formulation is briefly presented.  Section III contains our calculated
results and their discussions.  Finally, in the last section, a
summary is provided.

\section{Formulation and calculation}

The calculation of one-electron energy levels is based on the well
documented ${\bf k}\cdot{\bf p}$ approach.\cite{winkler}
For details about this method, e.g., the operator ordering, the
inclusion of a magnetic field, the influence of remote bands,
the influence of strain, and the modification due to heterojunction
interfaces, we refer to a partial list of publications and
references therein.\cite{bahder, burt, foreman, smith, zawadzki, rossler, chao}
In our calculation, the influence of strain is included and is
found to be important quantitatively.  The quantum well is assumed
to be parallel to the $xy$ plane, and the external magnetic field
is along the $z$ direction.  In our calculation, the axial
approximation is not used as mentioned in the introduction.
A plane wave and Landau level expansion scheme is used, and we
have checked carefully that there is no spurious solution in our
calculated electronic states.  The material parameters used in the
calculation can be found in the literatures\cite{chao, jancu}, and no other
adjustable parameter is introduced.

After obtaining the electronic energy levels, transition
energies can be easily calculated.  In order to know the nature of
a transition, we also calculate the corresponding optical
transition matrix elements between two involved states.\cite{ziman, yu}
Assuming the two states are denoted as $|1\rangle$ and $|2\rangle$,
we will calculate $\pi_x=|\langle 1|(p_x+eA_x/c)|2\rangle|^2$,
and $\pi_y=|\langle 1|(p_y+eA_y/c)|2\rangle|^2$.  Because of
the reduced symmetry in the InAs/GaSb quantum wells,
two matrix elements $\pi_x$ and $\pi_y$ are not identical, but
almost the same for the quantum wells studied in this paper.
In the calculation of the above matrix elements, one should take into
account the contribution from the Bloch basis states, as the
inter-subband optical transition is not fundamentally different
from the inter-band optical transition.

For simplicity, we will ignore the depolarization field
correction\cite{ando, kotthaus, tung}
due to the dynamic space charge effect in the quantum well.
We believe that this is a reasonable approximation, when the
magnetic field is not tilted, as the transitions involved have
their initial-states and final-states from the same subband.
The lack of detailed knowledge about the carrier distribution
also makes the investigation of this many-body effect difficult.
The exchange and excitonic effects due to the
electron-electron interaction\cite{exch} will also be neglected
for simplicity
in the present work, and will be studied in the future.

The InAs/AlSb/GaSb quantum wells studied in the present paper
have the following structure: the AlSb left-barrier, the InAs
layer, the thin AlSb layer, the GaSb layer, and finally the
AlSb right-barrier.
The growth direction of
the quantum wells is assumed to be [001].  The model quantum
well includes a thin layer of AlSb between the InAs and GaSb
layers.  The inserted thin AlSb layer can control the degree of
electron-hole hybridization.\cite{exp10, exp11}
The width of InAs layer will be varied, so that the relative positions
of the electron-like and the hole-like subbands can be changed.  The width
of inserted AlSb layer will also be varied.  The width of GaSb layer
will be fixed, so that the hole-like subbands are roughly unchanged
in energy.

\section{Results and discussions}

In Fig.1, the magnetic field dependence of Landau levels is shown
for two InAs/AlSb/GaSb quantum wells.  In the upper panel, the width
of InAs layer is $130$ angstrom, and in the lower panel, the width
of InAs layer is $170$ angstrom.  In both panels, the width of AlSb
layer is $10$ angstrom, and the width of GaSb layer is $50$ angstrom.
The quantum well structure information is explicitly shown in the
figure as QW:130/10/50 in the upper panel, and as QW:170/10/50 in
the lower panel.  The top of conduction band of InAs is taken as
the energy zero point.

The wave functions of some states are calculated.  By analyzing
the wave function of a state, one can easily determine that the
state is an electron-like (the conduction band contribution dominant)
state or it is a hole-like (the valence band contribution dominant)
state.  One can also determine the Landau level index of its
dominant contribution.  In Fig.1, when the largest component of
a state is from the conduction band, the energy of the state is colored red
for the spin-down state, and it is colored green for the
spin-up state.  From Fig.1, one can coarsely identify the fan
shape of the Landau levels of those electron-like states.  However,
the spin-up and spin-down electron-like Landau levels can no
longer form smooth fan-shape lines because of the electron-hole
hybridization.  One sees that usually a spin-up electron-like
state has a lower energy than the corresponding spin-down electron-like state.
This means that the $g$-factor is negative for the electron-like
states similar to the case of bulk InAs and the case of InAs
quantum wells.

In the energy window shown in Fig.1, the states whose energies
fall between the electron-like states are the hole-like states.
As the width of GaSb layer is fixed, the position of hole-like states
remains roughly unchanged, while the energies of electron-like
states decrease as the width of InAs layer increases.
This is expected, as the electron-like states are confined by
the outer AlSb barriers.  The slope of electron-like Landau levels
also changes as the InAs layer width varies.  This is due to
the non-parabolic band effect.

We find that the top energy levels, at about $0.11$ eV, of the
hole-like subband are zero index Landau levels.  At high magnetic
fields, the zero Landau levels from the electron-like subband
will grow in energy and intersect those hole-like zero Landau
levels.  There are small anti-crossing gaps but can not be seen with the
energy scale in Fig.1.  This zero Landau level anti-crossing
can be attributed to the influence of bulk inversion
asymmetry.\cite{hgte01, hgte02, hgte03}

Next, let us examine the Landau level transitions.  In Fig.2,
for the QW:130/10/50 quantum well (the width of InAs layer is $130$
angstrom, the width of AlSb layer is $10$ angstrom, and the
width of GaSb layer is $50$ angstrom), the energies of some Landau
level transitions are shown in the upper panel as a function of
the magnetic field strength.  The energy levels versus the
magnetic field are shown in the lower panel for the same quantum
well structure.

These transitions are from the states with energy lower
than $0.105$ eV to the states with energy higher than $0.105$ eV.
The selection for the Landau transitions is made because that the
transitions between the electron-like Landau levels must be included.
These transitions are most relevant to a magneto spectroscopy
experiment.\cite{exp11}  Some transitions from low lying states
to a few top hole-like Landau levels with low Landau level index
are also included, as the experimental samples usually contains
electrons as well as holes.\cite{exp10,exp11}
In the present work, we will make the same selection for the Landau
level transitions, thus one has a consistent view.

For a given magnetic field, the energy of a Landau level
transition is plotted in the upper panel as an open circle.
The size of the open circle is proportional to the corresponding
transition strength $\pi_x$.  At a fixed magnetic field, the
first $3$ most dominant transitions according to their $\pi_x$
values are colored.
It will be colored red if the corresponding initial-state is a
spin-down electron-like state.  It will be colored green if the
initial-state is a spin-up electron-like state.  If it can not
be colored red or green, it will be colored blue.
In the lower panel, the corresponding initial-state and final-state
of the Landau level transition are indicated with the same color.
The initial states are
marked as open circles, the final states are marked as open squares
or open diamonds.  Different symbol sizes are used in the lower
panel so that one can clearly see two transitions with the same
final-states.  We will focus on the high magnetic
field region, $B>15$T, as our calculated results can be presented
more clearly.

It must be pointed out that the strength of a mode observed in a
magneto spectroscopy experiment also depends on the involved sample carrier
density.  A transition will not be detected experimentally if
the initial-state is empty or the final-state is fully occupied.
We believe that the information in the way provided in Fig.2 should
be useful in analyzing an experimentally measured magneto-optical
spectrum.

In Fig.2, one sees that, at high magnetic fields, $B>15$T, the dominant
Landau level transitions can be viewed roughly as two spin-split
Landau level transitions with many electron-hole hybridization
induced splittings.  Two large hybridization induced splittings
occur at the magnetic field around $20$T.
For magnetic fields in the interval from $15$T to $20$T, the distance
between two dominant transitions is almost field independent first,
then decreases.  When the magnetic field is higher than $20$T, the
distance between two dominant transitions increases slightly as the
magnetic field increases.  From the lower panel of Fig.2, one sees
that, at high magnetic fields, the splittings in the Landau level
transitions are due to the electron-hole hybridization induced
initial-state splittings.

In Fig.2, one can observe electron-hole hybridization induced
multiple splittings on the dominant Landau level transition at a
very high magnetic field about $34$T.  This splitting occurs because
the $n=1$ hole-like Landau level has a non-monotonic magnetic field
dependence.  The $n=0$ electron-like Landau level shows anti-crossings with this $n=1$
hole-like Landau level at about $34$T.  It is clear that the size
of the splittings is much smaller than the splitting occurred at the
magnetic field about $20$T also due to the anti-crossing of the $n=0$
electron-like Landau level and the $n=1$ hole-like Landau level.
In a magneto-spectroscopy experiment, this splitting in the high
magnetic field may manifest as a resonance line-width broadening,
if the splitting is not fully resolved.

The degree of electron-hole hybridization can be tuned by changing
the width of inserted AlSb layer.  In Fig.3, the Landau level transition
energy versus the magnetic field strength is shown in the upper
panel for the QW:130/5/50 quantum well.  The AlSb layer width is reduced
to $5$ angstrom and the electron-hole hybridization hence becomes
stronger.  The lower panel of Fig.3 shows the energy levels versus
the magnetic field for the same quantum well structure.
By comparing the lower panels of Fig.2 and Fig.3, one can clearly
see the influence of this stronger electron-hole hybridization on
the energy levels.  At high magnetic fields, $B>15$T, the first $3$
most dominant transitions are
colored in the same way as that shown in Fig.2.  One sees that, at
high magnetic fields, the magnetic field dependence of the Landau
level transition energy shows a similar behavior as that displayed
in Fig.2.  As one expected, the hybridization induced splittings
around $20$T become larger, thus the Landau level transitions also show
larger splittings.  For the magnetic fields between $15$T and $20$T,
the distance between two dominant transitions also becomes larger.
One can still roughly view the dominant transitions at high magnetic
field, $B>15$T, as two spin-split Landau level transitions with many
electron-hole hybridization induced splittings.

Next, let us examine InAs/AlSb/GaSb quantum wells with a wider
InAs layer.  When the width of InAs layer increases, the energy
of an electron-like subband decreases, the gap between two
electron-like subbands becomes smaller.  For the QW:130/10/50
quantum well, there is one hole-like subband between two
electron-like subbands.  For the QW:170/10/50 quantum well, there
are two hole-like subbands between two electron-like subbands.
The Landau level transition
energies are shown versus the magnetic field in the upper panels
of Fig.4 for the QW:170/10/50 quantum well, and in Fig.5, for the
QW:170/7/50 quantum well.  The widthes of InAs and GaSb layers are
the same, only the widthes of inserted AlSb layers have a small
difference.  At high magnetic fields, $B>15$T, the transition
energies are displayed and
colored in the same way as that in Fig.2.  In the lower panels of
Fig.4 and Fig.5, the energy levels are shown as a function of the
magnetic field with the initial-states and final-states of Landau
level transitions indicated.

In Fig.4, when the magnetic field falls between $15$T and $20$T,
two dominant transitions almost coincide.  When the width of
inserted AlSb layer is reduced from $10$ to $7$ angstrom, one
can see clearly in Fig.5 that the distance between these two
dominant transitions becomes larger.
It should be pointed out that, one can still roughly view the
dominant transitions at high magnetic fields, $B>15$T, as two
spin-split Landau level transitions further split by many
more electron-hole hybridization induced splittings.

The tilting of the magnetic field is a technique often used in a
magneto-spectroscopy experiment.\cite{exp10, exp11}
One could obtain some information about the subband gap by tilting the field.
In Fig.6, the Landau level transition energies and energy levels
are shown as a function of the total magnetic field strength
for the QW:170/10/50 quantum well.  The magnetic field is tilted
$12.5$ degree along the [100] direction.  The transition energies
and Landau levels are displayed and colored in the same way
as that shown in Fig.2.  But, the first $4$ dominant transitions
are colored here.  As expected, one sees
that the tilted magnetic field induces a subband-Landau level
anti-crossing at about $35$T.  This anti-crossing field can be
reduced when one increases the width of InAs layer, thus reduces
the energy gap between the electron-like subbands.
By examining cases with different tilting
angles, we find that, the tilting of the magnetic field introduces
mainly a subband-Landau level coupling with the same spin.  One sees
that this subband-Landau level coupling affects the final-states more
strongly, and has a smaller effect on the initial-states.  For the
magnetic field lower than $30$T, the magnetic field dependence of
Landau level transitions is not strongly modified.  This can be
clearly seen by comparing Fig.6 to Fig.4.  This feature is also
observed in a recent experiment.\cite{exp11}  It is interesting
to note that the electron-hole hybridization induced splittings
can still be seen in the magnetic field region where the
subband-Landau levels are strongly coupled.
It should be pointed out that, the depolarization correction may
become important when the magnetic field is tilted.  However, one
needs the information about the carrier distribution in order to
take this effect into account.

%
%

\section{Summary}

In summary, the electronic structure of InAs/AlSb/GaSb quantum
wells embedded in AlSb barriers and in the presence of a
perpendicular magnetic field is studied theoretically within
the $14$-band ${\bf k}\cdot{\bf p}$ approach without making the
axial approximation.  The energies of some Landau level transitions
and their corresponding transition strengthes are calculated.
The magnetic field dependence of some dominant transitions is
shown with their corresponding initial-states and final-states
indicated.  This information should be useful in analyzing
experimentally measured magneto-optical spectra.  We find that,
at high magnetic fields, the dominant Landau level transitions
can be roughly viewed as two spin-split Landau level transitions
further split by many more electron-hole hybridization induced
splittings.

%
%

\begin{acknowledgments}

This work was partly supported NSF of China via
projects 61076092 and 61290303.

\end{acknowledgments}

%
%

%
%

\begin{figure}[ht]
\includegraphics[width=0.45\textwidth]{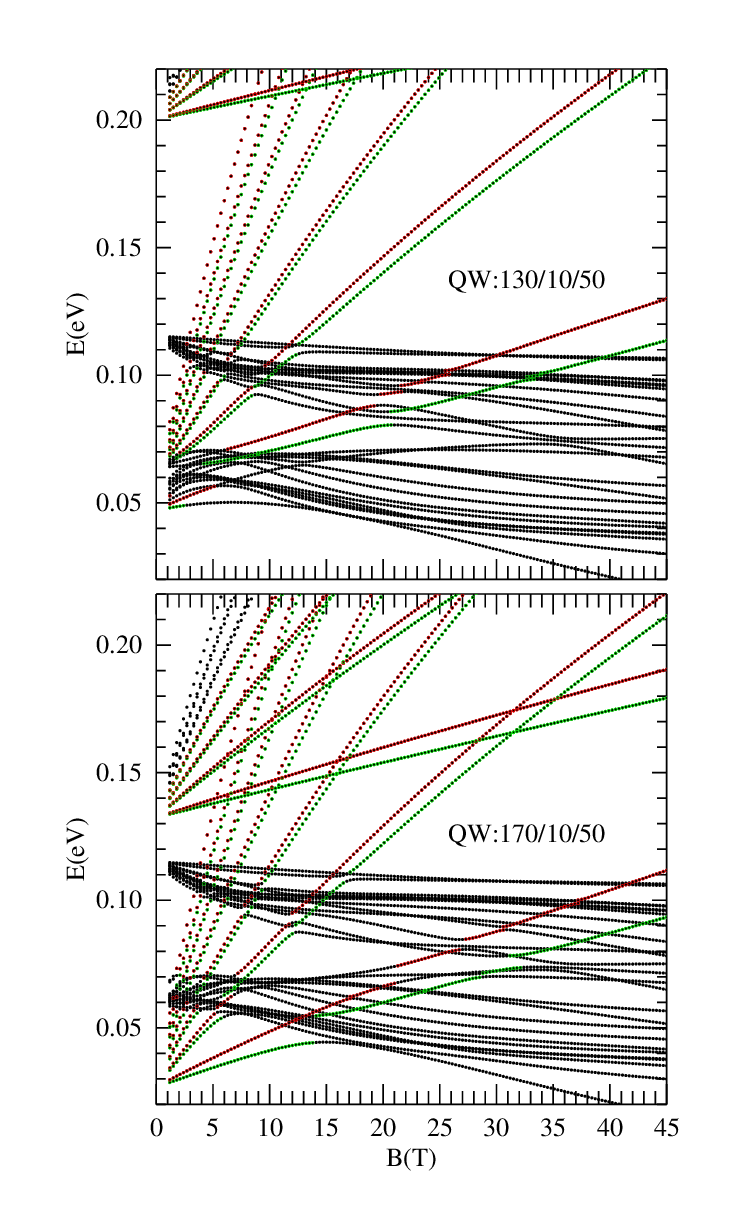}
\caption{(Color online)
The magnetic field dependence of Landau levels of InAs/AlSb/GaSb
quantum wells.  In the upper panel, the width of InAs layer is
$130$ angstrom, and in the lower panel, it is $170$ angstrom.
The width of AlSb layer is $10$ angstrom, and the width of GaSb
layer is $50$ angstrom.  When the largest component of a state is
from the conduction band, the dot symbol for the state energy is
colored red for the spin-down state, and is colored green for the
spin-up state.  The top of conduction band of InAs is taken as
the energy zero point.  The quantum well structure information is
denoted as QW:130/10/50 in the upper panel, and as QW:170/10/50
in the lower panel.
}
\end{figure}

\begin{figure}[ht]
\includegraphics[width=0.45\textwidth]{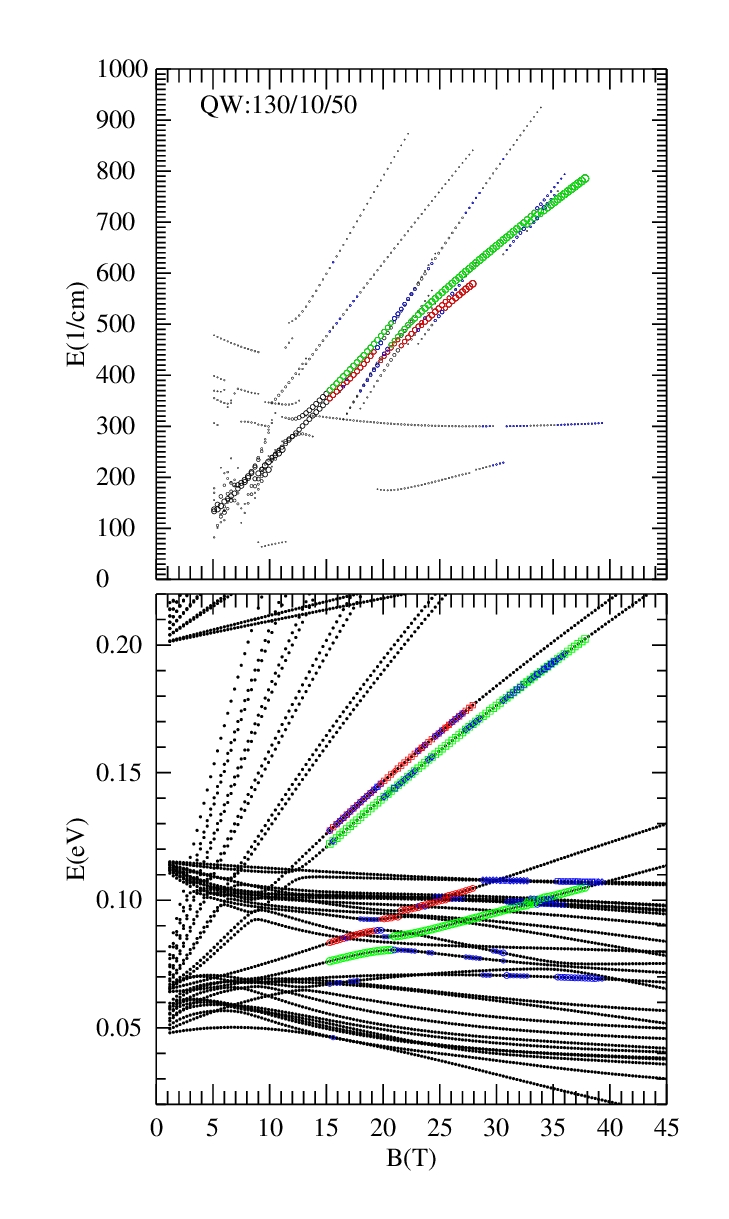}
\caption{(Color online)
The upper panel shows the energy of some Landau level transitions
(open circles) versus the magnetic field for the QW:130/10/50 quantum
well.  The widths of InAs, AlSb, and GaSb layers are $130$, $10$
and $50$ angstrom, respectively.  The lower panel shows the
energy levels versus the magnetic field for the same quantum
well structure.  These transitions are from states with energy
lower than $0.105$ eV to states with energy higher than $0.105$ eV.
The size of the open circles in the upper panel is proportional
to the corresponding transition strength $\pi_x$.  At a fixed
magnetic field, the first $3$ most dominant transitions according
to the corresponding $\pi_x$ values are colored red, green, and blue.
The corresponding initial-states and final-states of the Landau
level transitions are indicated in the lower panel with the same
color.  The initial-states are marked as open circles, the
final-states are marked as open squares or open diamonds.
}
\end{figure}

\begin{figure}[ht]
\includegraphics[width=0.45\textwidth]{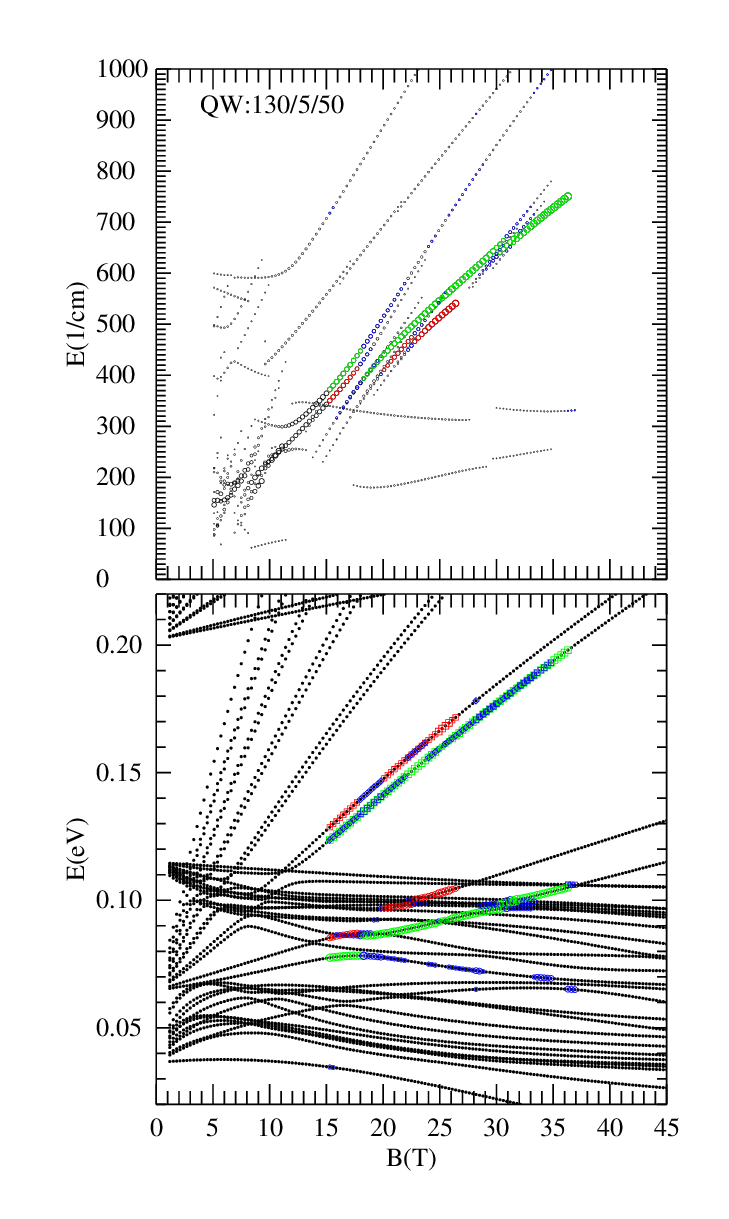}
\caption{(Color online)
The same as Fig.2, but for the QW:130/5/50 InAs/AlSb/GaSb
quantum well.
}
\end{figure}

\begin{figure}[ht]
\includegraphics[width=0.45\textwidth]{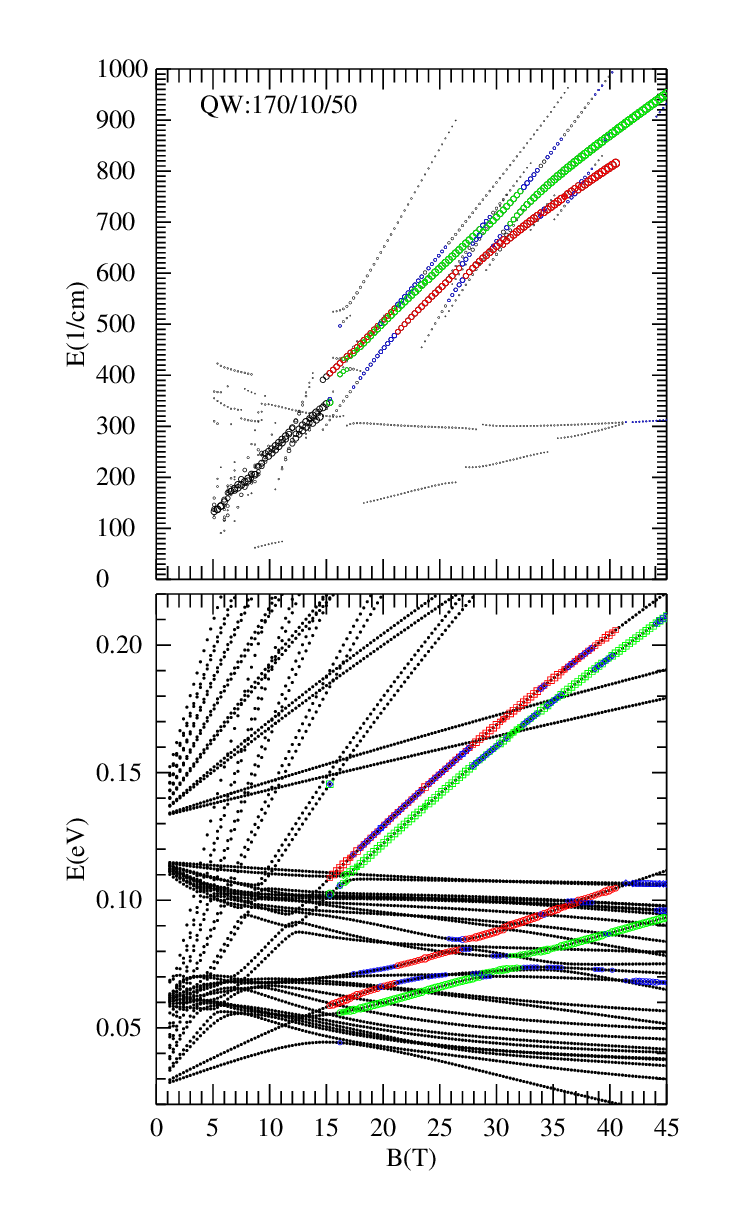}
\caption{(Color online)
The same as Fig.2, but for the QW:170/10/50 InAs/AlSb/GaSb
quantum well.
}
\end{figure}

\begin{figure}[ht]
\includegraphics[width=0.45\textwidth]{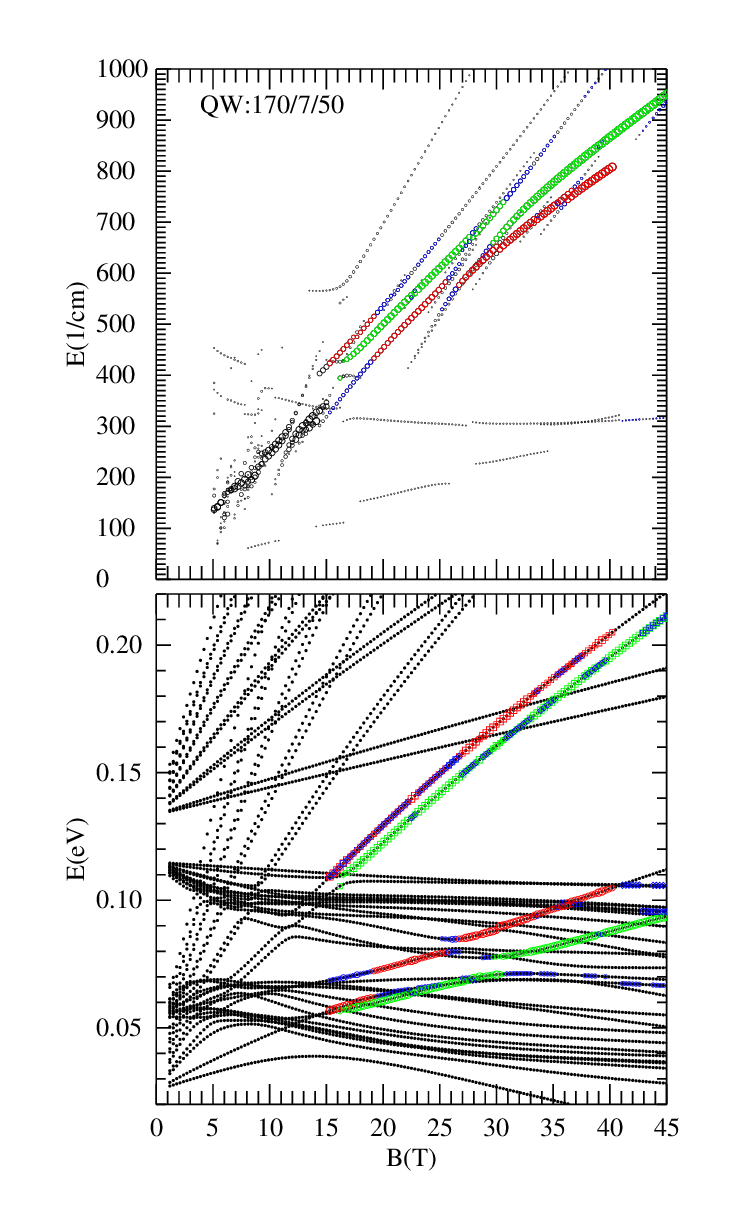}
\caption{(Color online)
The same as Fig.2, but for the QW:170/7/50 InAs/AlSb/GaSb
quantum well.
}
\end{figure}

\begin{figure}[ht]
\includegraphics[width=0.45\textwidth]{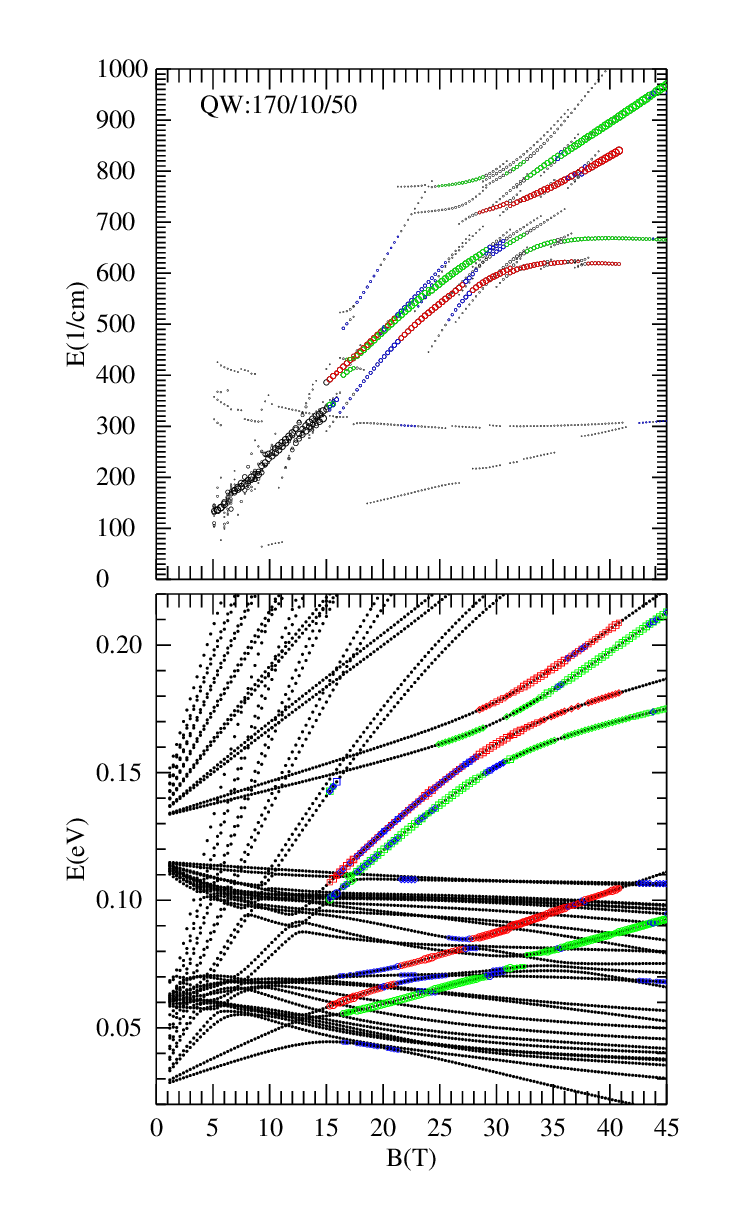}
\caption{(Color online)
The same as Fig.2, but for the QW:170/10/50 InAs/AlSb/GaSb
quantum well.
The magnetic field is tilted $12.5$ degree.
The first $4$ dominant transitions are colored.
}
\end{figure}

\end{document}